\documentclass[aps,preprintnumbers,nofootinbib,superscriptaddress]{revtex4}

\usepackage{graphicx}
\DeclareGraphicsExtensions{.pdf}

\usepackage{amsfonts,amssymb,amsmath}
\usepackage{amsthm}
\usepackage{color} 

\newcommand{\comment}[1]{}
\newcommand{\ket}[1]{\left |  #1 \right\rangle}

\theoremstyle{plain}

\theoremstyle{definition}
\newtheorem{definition}{Definition}

\begin{document}

\title{Device-Independent Relativistic Quantum Bit Commitment}

\author{Emily \surname{Adlam}} \affiliation{Centre for Quantum
  Information and Foundations, DAMTP, Centre for Mathematical
  Sciences, University of Cambridge, Wilberforce Road, Cambridge, CB3
  0WA, U.K.}  \author{Adrian \surname{Kent}} \affiliation{Centre for
  Quantum Information and Foundations, DAMTP, Centre for Mathematical
  Sciences, University of Cambridge, Wilberforce Road, Cambridge, CB3
  0WA, U.K.}  \affiliation{Perimeter Institute for Theoretical
  Physics, 31 Caroline Street North, Waterloo, ON N2L 2Y5, Canada.}

\date{April 2015}

\begin{abstract}
We examine the possibility of device-independent 
relativistic quantum bit commitment. 
We note the potential threat of {\it location attacks}, in
which the behaviour of untrusted devices used in relativistic quantum
cryptography depends on their space-time location.   
We describe relativistic quantum bit commitment 
schemes that are immune to these attacks, and
show that these schemes offer device-independent security against 
hypothetical post-quantum adversaries
subject only to the no-signalling principle.
We compare a relativistic classical bit commitment scheme
with similar features, and note some possible advantages of the 
quantum schemes.  
\end{abstract}

\maketitle

{\bf Introduction} \qquad  

This paper explores the scope for device-independent relativistic
quantum cryptographic protocols, focussing on the case of relativistic
quantum bit commitment.  There have been considerable advances
lately in both relativistic quantum 
cryptography
\cite{kentrel,kentrelfinite,nosummoning,qtasks,haydenmay,bcsummoning,otsummoning,colbeckkent,kenttaggingcrypto,malaney,buhrmanetal,kms,kentbccc,kentshort,bcmeasurement,crokekent,kthw,lunghietal,liuetal,practicalrbc,adlamkentdrbc} 
and device-independent quantum cryptography
\cite{mayersyao,bhk,agm,SGBMPA,abgmps,MRC,Masanes,HRW2,HR,MPA,BKP,bckone,colbeckphd,ck,Pironio,bcktwo,vv1,vv2}.   
This motivates our investigation of device-independent relativistic
protocols. 
Such protocols are not only potentially practically useful but also theoretically
interesting, because they add to our understanding
of the relationship between quantum nonlocality, cryptographic
security and the no-signalling principle. 
For example, asking whether certain
correlations are ``sufficiently nonlocal'' to implement a 
given task securely suggests useful measures of nonlocality. 
Determining which tasks can be implemented using quantum
or no-signalling resources is also a path to understanding the
properties of relativistic quantum information \cite{qtasks}. 
  
For tasks such as bit commitment, which involve two mistrustful
parties, device-independent protocols aim to guarantee security
to both parties even when neither of them trusts the quantum
devices used.  Fully device-independent protocols make no
assumption whatsoever about the mechanisms of the devices used, except
that (a) they are constrained by known physical laws, (b) they
can be contained within secure laboratories, whose resident
cryptographer controls all inward and outward information flows.
Thus both parties must allow for the possibility that any 
quantum devices they use may have been designed by the other party, who
may maliciously have included hidden features designed to give 
the appearance of normal functioning while actually compromising
security.    

One point we think particularly worth
highlighting is that requiring full device-independence is an 
even stronger constraint in relativistic quantum cryptography
than in standard quantum cryptography.   Relativistic cryptographic
protocols require specific configurations of agents and devices
in space-time, arranged to ensure that the agents cannot exchange
data during the protocol (or critical parts thereof).     
Thus if two separated agents are using devices that are supposed to
produce appropriately correlated outputs, they may be unable to check 
this behaviour during critical parts of the protocol.  
Even if they perform tests of the devices before the protocol, 
they remain vulnerable to space-time {\it location attacks}.
That is, devices that are able to track their own space and time
coordinates may be programmed to reproduce the expected  
behaviour when they are originally tested but not when they are implementing
a critical part of the protocol at specified space-time locations.  
For example, if a particularly
critical bit commitment is going to be initiated at noon GMT by
devices in Cambridge and Auckland, one of Alice's  
devices might be programmed so that it usually behaves as Alice
expects but, if and only if it is in Cambridge
at noon, it gives her an output that should encrypt her committed bit 
but actually immediately reveals it to Bob's Cambridge agent.
These attacks are potentially problematic, because they defeat some testing
strategies that work well in non-relativistic device-independent
quantum cryptography.  Nonetheless, the protocols we describe resist
them.

From the practical point of view, one would also like to 
understand whether device-independent relativistic 
quantum bit commitment protocols have
any advantage over relativistic classical bit commitment 
protocols.  Device-independent protocols aim to ensure security 
despite a lack of trust in quantum devices, but still require
trusted classical computing, storage and communication devices.
Classical bit commitment protocols also require trusted classical
devices, but require no quantum devices.  It is well known that there 
are unconditionally secure classical relativistic bit
commitment protocols \cite{kentrel,kentrelfinite}. 
While these involve somewhat different configurations of
agents in space-time from the device-independent protocols
we consider here, we describe below another 
unconditionally secure classical relativistic bit commitment
protocol that has the same configuration of agents as our
device-independent protocols.   Our  
device-independent quantum schemes thus cannot offer
any advantage on this point.  
However, relativistic quantum bit commitment protocols do have
some subtler advantages in some scenarios \cite{lunghietal,adlamkentdrbc},
which one can seek to maintain in device-independent versions.     
We discuss these further below.  

{\bf Device-Independent Relativistic Quantum Bit Commitment Protocols} \qquad 

We begin by considering device-independent versions of the two
entanglement-based relativistic bit commitment protocols described in
Ref. \cite{adlamkentdrbc}.  The protocols of Ref. \cite{adlamkentdrbc} are
unconditionally secure given the validity of quantum mechanics and
special relativity, assuming that the participating agents' trust the
functionality of their quantum preparation and measuring devices, up
to limited losses and errors.  Here we describe variations whose
security does not require trusted devices.  These protocols
can be proven secure assuming only the 
cryptographic no-signalling principle \cite{bhk}:  

\begin{definition} 

  \textbf{No-signalling principle:}  Let $D_i$ (for $1 \leq i \leq N$)
  be devices in separate secure laboratories at points $P_i$,
  which accept inputs $A_i$ and produce outputs $X_i$. 
  Let $J$ be any subset of $ \{ 1 \ldots N \}$, let $\bar{J}$ be the
complementary subset, and let $X_J$ be the random variable $\{ X_j \}_{j \in J}$.   
 Then the  conditional probability distributions for
  the outcomes satisfy $p(X_J | A_1 , \ldots , A_N ) = p( X_J | A_J
  )$.  
 \footnote{If both inputs
and outputs take place at the same points $P_i$, then ensuring that
these points are spacelike separated is a sufficient condition.
It is not necessary, however.   
The cryptographic assumption of secure laboratories implies that
the devices cannot communicate input or output information in any way
through the laboratory walls, whether or not they are spacelike separated.} 

\end{definition}

Since our security proofs make no assumptions about the properties of
preparation or measurement devices, these protocols offer
device-independent security. Moreoever, the proofs do not assume the
validity of quantum mechanics, and therefore these protocols are also
secure against adversaries who can exploit hypothetical post-quantum
non-signalling theories.

{\bf Bit commitment}\qquad We employ the definitions set out in
Ref. \cite{adlamkentdrbc} for bit commitment and relativistic bit commitment,
and refer to the discussion therein for further commentary.   

In the protocols we consider below, there are two unveiling agents
$A_0$ and $A_1$, whose actions are spacelike separated from each other
and from those of $A_c$. 
The probability of a successful unveiling of bit value $i$ depends
only on the actions of agent $A_i$. 
We will say that such a protocol provides unconditional, device-independent
security against Alice if any only if for any collective strategy $S$ which is
possible according to the no-signalling principle, $p_0 (S) + p_1 (S)
< 1 + \epsilon (N)$ and $\epsilon (N ) \rightarrow 0$ as $N
\rightarrow \infty$, where $N$ is a variable security parameter of the
protocol, and $p_0 (S)$ and $p_1 (S)$ are the probabilities that, by
following strategy $S$, Alice and her agents persuade Bob that they
have validly committed and unveiled $0$ or $1$ respectively according to the rules
of the protocol. Note that a collective strategy may be fixed by Alice before the
protocol, or Alice's agents responsible for unveiling $0$ and $1$ may
independently choose their strategies after the commitment time,
possibly conditioned on events in the past lightcone of their
verification point but not of the commitment point. We subsume the
latter possibility under the former by allowing strategy $S$ to
include steps in which agents make strategic choices with
probabilities conditional on certain external events, with those
events themselves now explicitly included in the description of 
strategy $S$. Any strategy whereupon the conditional
probabilities for these choices are nontrivial may be written as a
convex combination of deterministic strategies, so no probabilistic
strategy can have greater success probability than the most successful
deterministic strategy.

We say that a relativistic protocol is unconditionally secure against Bob if, for 
any strategy Bob's agents may follow that is consistent with the
no-signalling principle, if Alice's agents choose not to 
unveil, then the probability of any of Bob's agents correctly
guessing the committed bit at any point in space-time
is bounded by $ 1/2 + \epsilon' (N)$, where  
$\epsilon' (N ) \rightarrow 0$ {as} $N \rightarrow \infty$.  
It follows from this definition, by the no-signalling principle,
that when Alice does choose to unveil,
Bob cannot guess Alice’s commitment anywhere that does not lie in the
future lightcone of the unveiling points.

In the protocols we consider below, Alice has one committing agent,
$A_c$, and two unveiling agents, $A_0$ and $A_1$, who can 
unveil a valid commitment to $b=0$ and $1$
respectively.  
An additional security criterion may be required for such protocols:
that if $A_c$ does not make a valid 
commitment to bit value $b$, $A_b$ follows the unveiling
protocol and $A_{\bar{b}}$ does not, then Bob's agents, at any point
in space-time, should gain no information about whether $A_c$
committed to bit value $\bar{b}$ or declined to make a valid
commitment.   Our protocols also satisfy this criterion. 

As usual in quantum cryptography, we initially present our protocols in an 
idealized form assuming perfect quantum state preparations,
transmissions, measurements and computations.  
However, the protocols are 
tolerant to noise and errors, as we discuss later. 

{\bf Space-time and communications} \qquad 
We also make standard idealizations about the background geometry and 
signalling speed.   We suppose that space-time is Minkowski
and that Alice and Bob each have agents in secure
laboratories infinitesimally separated from the points $P$, $Q_0$ and
$Q_1$, that signals are sent at precisely light speed, and 
all information processing is instantaneous.  
Again, these assumptions can be relaxed.  The protocols remain  
secure in realistic implementations with finite separations and 
near light speed communication.  If these corrections are
small, the only significant effect is that 
Bob is guaranteed that Alice's commitment
is binding from some point $P'$ in the near causal future of $P$,
rather than from $P$ itself \cite{bcsummoning}.  
Allowing for small deviations from Minkowski geometry also  
requires small corrections to the geometry when stating the security
guarantees, but does not essentially affect security beyond that \cite{kentrelfinite}.  

{\bf Geometry} \qquad Alice and Bob agree on a space-time point $P$ and
an inertial set of coordinates $(x,y,z,t)$ for Minkowski space with
$P$ as the origin.  We focus here on the simplest case in which there
are two possible unveiling points $Q_0$ and $Q_1$; the protocols straightforwardly extend
to versions with $N$ unveiling points committing
  ${\rm log} (N)$ bits. 
Alice and Bob each have agents, who
during the protocol are separated in secure laboratories, adjacent to
each of the points $P$, $Q_0$, $Q_1$.  To simplify we
take the distances from these labs to the relevant points as
negligible.  Let the agents adjacent to $P$ be $A_c$ and $B_c$, and those adjacent to
$Q_i$ be $A_i$ and $B_i$.

For the two protocols we describe here, the unveiling points must be
spacelike separated from the commitment
point. Although it is not necessary for much of our discussion,
we assume that $Q_0$ and $Q_1$ have positive time coordinates in the
given frame, and therefore these protocols  are {\it fixed frame positive duration} (FFPD) 
relativistic bit commitments in the terminology of Ref. \cite{adlamkentdrbc}.  

We will also suppose that Bob has an unlimited number of additional
agents which he may position at various points between the the
commitment point $P$ and the unveiling points $Q_0$ and $Q_1$. However
Bob needs these agents only if he is concerned about verifying Alice's
commitment as soon as possible; if the situation is not time-sensitive
the verification steps described below could be carried out either by
$B_c$ or $B_i$ at a slightly later time.

For definiteness we describe a protocol in which Alice and her agents
exchange qubits by physical transportation in the preparation
phase.   They may alternatively employ a secure quantum channel without
materially altering the security of the protocol.  Note however that 
assuming a physically secure quantum channel requires that the physical path of the
channel lies within Alice's secure laboratories. 
If one instead supposes they use quantum teleportation, then to make 
a fair comparison of a fully device-independent scheme against 
alternatives, one would need to discuss further the trust 
Alice places in her quantum teleportation devices
and the encryption or other resources she may consume.    
  
{\bf Notation} For the security proofs we employ a binary notation,
denoting measurement directions $X, Y$ by $0, 1$, measurement
directions $X', Y'$ also by $0, 1$, and measurement outcomes $+1,
-1$ also by $0, 1$. A single Bell experiment is then described by a
choice of two measurement settings $(x, y) \in \{ 0, 1 \}$ and two
measurement results $(O, s) \in \{ 0, 1 \}$. 
We are interested in the {\it CHSH game score} $R(t \oplus s = xy)$,
which is the number of experiments in the set for which the settings $(x, y)$ and
outcomes $(t, s)$ satisfy the condition $t \oplus s = xy$. 
In the case when the settings are random and independent,
the CHSH game score is asymptotically related to the average value of
the CHSH observable $(XX' + XY' + YX' - YY')$: for a set of $N$ 
experiments, the latter expression is $\frac{4}{N} (2 R(t \oplus s =
xy) - N)$ to leading order in $N$.  
However, we do not assume independent random settings 
in all of our protocols.   
\vfill

\underline{{\bf{CHSH 1: \bf{CHSH test protocol with fixed directions}}  
}} \qquad 
 
{\bf Preparation 1.} Alice and Bob agree on a set of four directions
$X, Y, X', Y'$ such that all four directions lie in a single plane,
$X$ is orthogonal to $Y$, $X'$ is orthogonal to $Y'$, $X'$ is
separated from $X$ by $\frac{\pi}{4}$ and $Y'$ is separated from $Y$
by $\frac{\pi}{4}$. They also agree on a bit string $L^0$ with bitwise
complements $L^1$. 

{\bf 2.} $A_c$ instructs her devices to prepare $2N$ Bell pairs $(W_a^i, W_b^i)$ in the
singlet state $\Psi^-$, then randomly draws an integer $x \in \{ 0,
1\}$. If $x = 0$ she gives the particles $[ W_b^1, ... W_b^N]$ to an
agent $A_0$ and the remaining particles $[ W_b^{N + 1}, ... W_b^{2N}]$ to
an agent $A_1$; if $x = 1$ she instead gives the particles $[ W_b^{N + 1},
... , W_b^{2N}]$ to $A_0$ and the particles $[ W_b^1, ... , W_b^N]$ to
$A_1$. $A_0$ and $A_1$ then travel to points $Q_{0}$ and $Q_{1}$ both
at a distance $d$ away from $A_c$. 

We assume that $A_c$, $A_0$ and
$A_1$ have secure laboratories that protect their qubits, so $B_c$
cannot interfere with them in any way after the initial preparation.
In particular, $A_0$ and $A_1$ travel within secure laboratories.
We also assume that the first and second sets of states (i.e. the
purported half-singlet states labelled $a$ and $b$ respectively)
are securely separated in sub-laboratories
before $A_c$ chooses the value of $x$.
Thus, even if the $a$ states are actually malicious devices 
operating within the laboratory, they obtain no information about 
the value of $x$ or the destinations of the $b$ states.    

{\bf Commitment} At the designated commitment point $P$, $B_c$ gives
$A_c$ a bit string $L$ drawn uniformly at random from
$\mathbb{Z}^{\otimes N}_2$.  If $A_c$ wants to commit to $0$, she
instructs her devices to measure 
each qubit $W^{j + xN}_a \in [W^{1 + xN}_a, ... , W^{N +
  xN}_a] $ in direction $(X)^{L_j} (Y)^{1 - L_j}$. If $A_c$ wants
to commit to $1$ she instead instructs her devices 
to measure each qubit $W^{N + j - xN} \in
[W^{N + 1- xN}_a, ... , W^{2N - xN}_a] $ in direction $(X)^{L_j }
(Y)^{1 - L_j}$. In either case $A_c$ immediately broadcasts her
outcomes.

{\bf Unveiling} For $i \in \{0, 1 \}$: 

At the point $Q_i$, if agent $A_i$ believes Alice
wishes to unveil, she measures each qubit {$W_a^{j + x N + i N - 2 i x N} \in [W^{1 + x N + i N - 2 i x N} _a
...W^{N + x N + i N - 2 i x N}_a] $ }in direction
$(X')^{L^i_j} (Y')^{1- L^i_j }$. 
She then broadcasts her measurement outcomes.

Note that in principle the agents $A_c$, $A_0$ and $A_1$ may make
these decisions independently.   If {Alice} wishes to coordinate them and ensure that
all or none unveil, she needs to give them instructions in
advance.  These instructions could depend on separate events in the
past light cones of their unveiling decision points, if Alice believes
these events will be correlated.

{\bf Verification} Bob's agents between $P$ and $Q_i$ wait for the
data from $A_c$ and $A_i$. As soon as one agent receives both sets of
data, he calculates the CHSH game score. 
If the result is greater than $ N ( ( 2 + \sqrt{2} )/4 - \xi )$
he accepts that Alice made a valid commitment to bit value $i$.
Here $\xi$ is some predetermined small security parameter,
chosen such that $\xi \gg N^{-1/2}$; this ensures that the probability 
that $N$ correctly prepared and measured singlets will fail
Bob's test is suitably small.   

\vspace{2mm} {\bf Security against Alice}  $A_0$, $A_c$ and $A_1$
announce their outcomes at spacelike separation from one another
during the protocol.  They may pre-agree a collective strategy $S$, which
may rely on shared quantum or (hypothetical) post-quantum
no-signalling resources, but cannot signal to one another during the protocol.  
We are interested in bounding the probability that $A_0$ and $A_1$
both succeed in passing Bob's tests for a valid commitment of 
$0$ and $1$ respectively.   

Let $p_0 (S)$ and
$p_1 (S)$ be the probabilities that when Alice employs strategy $S$,
she and her agents convince Bob that they have have unveiled $0$ or
$1$ respectively according to the rules of the protocol above.  Note
that these rules allow for the possibility that Bob accepts valid
commitments to both $0$ and $1$ if both tests are passed.
Let $p_0 (S) + p_1 (S) = 1 + \epsilon^1_S(N)$.

Hence for any strategy $S$, $\epsilon^1_S(N) \leq \sum_{L}
\frac{p^S_{L,L_0}}{2^{N}}$ where $p^S_{L, L_0}$ is the probability
that in this protocol (defined by $L_0$),
when Alice employs strategy $S$ and Bob chooses the bit string $ L $, 
Alice and her agents produce three sets of outcomes $O, O^0$ and
$O^1$ such that: 

\begin{align} d(O^0 \oplus O, L^0 L) \leq N(\frac{1}{2} -
  \frac{1}{2\sqrt{2}} + \xi ) \label{equ1} \end{align}

\begin{align} d(O^1 \oplus O, L^1 L) \leq N(\frac{1}{2} -
  \frac{1}{2\sqrt{2}} + \xi ) \label{equ2}
\end{align} 

Here $d(x,y)$ denotes the Hamming distance
between the bit strings $x$ and $y$, which is the number of
positions on which the strings differ, $O^i \oplus O$ is the string
given by element-wise modular addition {of $O^i$ and $O$}, and $L^i L$ is the string given
by element-wise multiplication {of $L^i$ and $L$}. 

These equations can be simultaneously satisfied only if $d(O^0 \oplus
O^1 , L(L^0 \oplus L^1)) \leq N( 1 - \frac{1}{\sqrt{2}} +  2  \xi )$.  
Since $L^0$ is the bitwise complement of $L^1$, this
implies that $d(O^0 \oplus O^1 , L) \leq N( 1 -
\frac{1}{\sqrt{2}} +  2 \xi )$.  
 
Thus for given $O^0 \oplus O^1$, equations \ref{equ1} and \ref{equ2}
may be satisfied simultaneously only if $L$ lies within the Hamming
ball $H$ of radius $r = N(1 - \frac{1}{\sqrt{2}} + 2 \xi )$
centered on $O^0 \oplus O^1$.  For $r \leq \frac{N}{2}$, 
the volume of this ball is less than or
equal to $2^{N H(r/N)}$ where $H(x) = -x \log(x ) - (1 - x )\log(1 - x
)$ \cite{Rudra}.
 
$H(x) \leq 1$ with equality if and only if $x = \frac{1}{2}$, so for
any $\xi < \frac{1}{2 \sqrt{2}} - \frac{1}{4}$ we have $H(r/N) < 1$.

By the no-signalling principle, the probability distribution for $O^0
\oplus O^1$ is independent of $L$.  Thus since $L$ is chosen uniformly at
random from $\mathbb{Z}^{\otimes N}_2$, we must have $\epsilon^1_S(N)
\leq 2^{-N (1 - {H(r/N)})}$, which for $H(r/N) < 1$ goes to zero as N goes to infinity.

\vspace{2mm} 
{\bf Security against Bob} \qquad 

Alice's devices may perhaps have been
designed by Bob in an attempt to cheat the protocol.  
If the devices do something other than performing CHSH measurements 
on a shared quantum singlet, $A_c$'s outputs  may give $B_c$
information about whether $A_c$ measured the
qubits $[W^{1}_a ...W^{N}_a] $ or $[W^{N + 1}_a ...W^{2N}_a] $.
This may allow $B_c$ to update his prior values (which are
originally equiprobable) for $P(b | x )$, 
the conditional probability of the committed bit value $b$ given
the value of $A_c$'s randomly chosen bit $x$.
However, by assumption, $x$ is random and kept secret throughout
the protocol.   Before receiving unveiling data, Bob's estimate
of $P( b) =  \frac{1}{2} ( P (b | 0 ) + P(b| 1) )$ thus remains
unaltered.    

Security against Bob thus relies on Alice being able to generate one
unconditionally secure random bit per committed bit, and 
to store this bit securely in $A_c$'s laboratory during the protocol.   

\vspace{2mm} 

\underline{\bf{CHSH2: CHSH protocol with secret complementary bit
    strings}} CHSH1 may straightforwardly be varied so
that Bob keeps the bit strings $L^0$ and $L^1$ secret from Alice until
the points $Q^0$ and $Q^1$:

{\bf Preparation 1.} $B_c$ chooses a length $N$ bit string $L^0$ drawn
uniformly at randomly from $\mathbb{Z}^{\otimes N}_2$. He communicates
string $L^0$ to $B_0$ and its bitwise complement $L^1$ to $B_1$. 

The remaining steps are identical to those for CHSH1. 

\vspace{3mm} {\bf Security against Alice:} The only change in this
protocol is that Alice is given less information, which can only
decrease her probability of cheating successfully. Hence the security
of this protocol against Alice follows immediately from the security of
CHSH1 against Alice.

{\bf Security against Bob} \qquad The proof of security against Bob is
the same as for CHSH1.

\vspace{5mm}

\underline{{\bf{CHSH3:} \bf CHSH test protocol with full randomisation}}

\qquad CHSH1 may also be varied so that the measurement directions
for $A_0$ and $A_1$ are not fixed in advance.  Instead, each $B_i$
randomly selects a set of $N$ measurement directions to be used by
$A_i$, so the string $L^1$ is no longer guaranteed to be the bitwise
complement of $L^0$.

The steps in this protocol are identical to those for CHSH1 
except for the unveiling stage:
 
{\bf Unveiling} For $i \in \{0, 1 \}$: 

At the point $Q_i$, $B_i$ gives $A_i$ a bit string $L^i$ drawn
uniformly at randomly from $\mathbb{Z}^{\otimes N}_2 $.
If agent
$A_i$ believes Alice wishes to unveil, she
immediately measures each qubit $W_a^{j + x N + i N - 2 i x N} \in [W^{1 + x N + i N - 2 i x N} _a
...W^{N + x N + i N - 2 i x N}_a] $  in direction $(X')^{L^i_j} (Y')^{1 - L^i_j }$.  
She then broadcasts her measurement outcomes.

{\bf Security against Alice}

As before, for any strategy $S$, let $p_0 (S)$ and $p_1 (S)$ be the
probabilities that when Alice employs the strategy, Alice's agents
convince Bob that they have have validly unveiled $0$ or $1$
respectively according to the rules of the protocol, and suppose $p_0
(S)+ p_1 (S) = 1 + \epsilon^2_{S}(N)$.  

By the no-signalling principle, $A_0$'s success probability is
independent of the choice of $L_1$, and $A_1$'s success probability
is independent of the choice of $L_0$. 
Hence we may calculate $p_0 (S)$ and $p_1 (S)$ 
assuming that $L_0$ and $L_1$ are
bitwise complements.  
We thus obtain the same bounds on $p_0 (S) + p_1 (S)$ as for
the first two protocols.  

{\bf Security against Bob} \qquad The proof of security against Bob is
the same as for CHSH1.

\vspace{5mm}

\underline{{\bf{Extensions}} }\qquad 

{\bf Declining to commit} \qquad 

In these relativistic bit commitment schemes, $A_c$ decides whether or
not to make a commitment.  If she makes a decision at the point $P$,
she cannot communicate it to the $A_i$ before they make decisions 
about whether or not to unveil.   Bob will learn whether or not
a commitment was made and validly unveiled if he receives unveiling
data.  However, Alice may perhaps not wish Bob to learn whether
or not $A_c$ chose to make a valid commitment in the cases
when the $A_i$ choose not to unveil.     

A subtle potential drawback of the protocols described above is that the
security proofs do not guarantee that Bob will be unable to tell
whether or not $A_c$ has chosen to make a commitment. 
If Alice's devices actually implement CHSH measurements on singlets,
then $A_c$ could decline to commit simply by returning a random
string of $N$ bits to $B_c$.   
However, Bob may have designed Alice's preparation devices.  He thus 
may be able to distinguish between randomly chosen bit strings and
strings produced by measurements on Alice's Bell pairs, and may therefore
may be able to detect if $A_c$ tries this tactic to refrain from making a
commitment.

However, in scenarios where these feature is problematic, our protocols can straightforwardly be altered to eliminate it. 
Alice and Bob may simultaneously
perform two runs of CHSH1, with the rule that
Alice commits to bit value $b$
in both protocols if she wishes to produce a valid commitment to $b$,
but commits to different values in each protocol if she wishes to
refrain from commitment. Bob will then accept that Alice has made a
valid commitment to bit value $b$ only if she unveils valid
commitments to bit value $b$ for both protocols.  
CHSH2 or CHSH3 may similarly be duplicated in this way.  

Bob obtains no information about the values of $x$ used in 
the two protocols unless Alice unveils.  He thus cannot tell whether
or not she made a valid commitment if she declines to unveil. 

{\bf Errors and losses} \qquad 

We now consider how to allow for the
possibility of small errors in preparations, transmissions and measurements
of quantum states.
We consider an error model which allows for errors and losses in the preparation,
transmission and measurement of singlet states. 
We suppose that any such errors occur randomly and independently, and
that the combined rate of errors and losses is bounded by a 
parameter $\delta$, which we assume is small. 
In this model, the expectation values of the distances $d(O^i \oplus O, L^i L)$
are altered by no more than $\delta N$, with variance no larger than 
$ \delta^{1/2} N^{1/2}$.   

Bob's verification tests involve a security parameter $\xi$, and the
security proof for CHSH1 holds for 
\begin{equation}\label{xirange}
\xi < \frac{1}{2 \sqrt{2}} - \frac{1}{4} \, . 
\end{equation} 
In particular, if $\delta$ is agreed to be the maximum tolerable rate
of errors and losses, and $\delta$ is small, Bob can use a value $\xi
> \delta$ in his verification tests,  consistent with (\ref{xirange}), 
and still ensure security against Alice.  
  
The security proofs of CHSH2 and CHSH3 do not 
depend on additional assumptions about
errors and losses, and so similarly remain secure for small $\delta$. 
Thus we can ensure that both protocols are secure against
Alice provided the expected rate of errors and losses is small. 

\vspace{2mm} The proofs of security against Bob do not depend on any
assumptions about errors or losses, and therefore no alteration is
required for these proofs.

\vspace{3mm}  

{\bf Discussion}

{\bf Device memory attacks} \qquad 

A quite general concern in device-independent quantum cryptography is
that maliciously designed devices may keep records of their inputs and
outputs and may make their future outputs depend on these data
\cite{bckone}.   Our protocols are secure against these attacks for a 
single bit commitment.   However, if Alice reuses her devices for
a sequence of commitments, our protocols do not protect her from
the possibility that data released (for example to $B_c$) in a 
later commitment may give information about earlier commitments. 
This is not a concern if all her commitments are in any case 
later unveiled, so long as the devices are not reused until
points in the causal future of points at which Bob 
can receive both the commitment and unveiling data.
Similarly, it would not be a concern if her commitment data loses any value
in the causal future of certain spacetime points (call them {\it
expiration points}), whether or not it was unveiled, so long as the 
devices are not reused until points in 
the causal future of an expiration point.   However, in many common
scenarios,
some commitment data is unveiled and some should be kept
secret indefinitely.  Our protocols do not guarantee security
for device reuse against completely general attacks in such
scenarios.     

However, our protocols do guarantee security for device reuse 
against memory attacks, so long as the devices are not also 
able to carry out location attacks, and so long as the $A_i$'s
decision whether to unveil is coordinated. 
The reason is that, if neither $A_0$ and $A_1$ 
accept any device input or supply any output unless the bit is
to be unveiled, and if the devices are not sensitive to their
location, then the devices can carry no 
information correlated with $x$ (and hence nor with $b$) after a 
protocol without unveiling.  They thus have no ability to 
leak information about $b$ in future protocols.  

{\bf Other security assumptions} \qquad 

As with the protocols of 
Refs. \cite{kentrelfinite,bcsummoning,bcmeasurement}, the present
protocols can be chained together in sequence, allowing longer term
bit commitments and flexibility in the relation between the commitment
and unveiling sites (in particular, they need not be lightlike
separated).  Full security and efficiency analyses for these chained
protocols remain tasks for future work.

We note that the unconditional, device-independent security of both
protocols depend on the assumption that Bob has access to a device
independent method of generating random numbers which is secure in the
sense that Alice cannot predict its output in advance.  
There exists no device independent method of generating random numbers
with no random seed, but there do exist device independent methods of
randomness expansion \cite{colbeckphd,ck,Pironio,vv1}, which can create
a long random string from some small random seed. Since the random
seed used in these protocols can be of a low quality of randomness
(i.e. the expansion still works if the `random' input is partially
correlated with variables which may be known to the adversarial party
\cite{colbeckrenner}), it could be obtained by some classical quasi-random
process which Alice is unlikely to be able to predict or control
perfectly, such as a coin flip.  Therefore device-independent
randomness expansion suffices for effectively device-independent
security of these protocols. 

The CHSH2 protocol requires Bob to generate $2N$ random strings and
the CHSH3 protocol requires him to generate $3N$ random strings, in
contrast with only one string in CHSH1.  However there may be
contexts in which CHSH2 and CHSH3 are preferred - for example, if Bob
wishes to control the time of Alice's measurements, which might be
particularly important for chained protocols. In the CHSH1 protocol,
Alice may make the measurements whose results she unveils at $Q_0$ and
$Q_1$ at any time, even before the start of the protocol, but in 
CHSH2 and CHSH3 she must wait until Bob tells her which measurement directions
to use. In such contexts CHSH3 offers greater security to Bob,
because in CHSH2, $L_1$ must be the bitwise complement of $L_0$ and
therefore it follows from the no-signalling principle that $B_0$ and
$B_1$ must agree on these strings at some time in the causal past of
$P$. This is a potential security weakness, since if Alice can find
out the strings $L_0$ and $L_1$ in advance of the commitment, then
$A_0$ and $A_1$ can perform their measurements earlier than Bob
expects.

Note that, like all technologically unconstrained quantum bit commitment
protocols\cite{kentbccc,kentshort}, our protocols do not 
prevent Alice from committing to a quantum superposition of 
bits.   She can simply input a superposition $\alpha \ket{0} +
\beta \ket{1}$ into a quantum computer programmed to implement
the two relevant quantum measurement interactions for inputs 
$\ket{0}$ and $\ket{1}$ and to send two copies of the quantum
outcome data towards $Q_0$ and $Q_1$, and keep all the data at
the quantum level until (if) she chooses to unveil.   
This gives her no advantage in stand-alone applications of bit
commitment, for example for making a secret prediction: it 
does, however, mean that one cannot assume that in a task
involving bit commitment subprotocols, any unopened
bit commitments necessarily had definite classical bit values,
even if all unveiled bit commitments produced valid classical
unveilings.

In comparing relativistic bit commitment
protocols, one needs to consider the configuration of agents in
space-time as well as the resources required. 
To make a precise comparison, we first describe a classical
relativistic bit commitment protocol with the same configuration of agents
as the schemes above. 
\vskip 10pt 

\underline{{\bf Random code classical bit commitment protocol (RCCBC)}} \qquad

{\bf Preparation 1.} Alice and Bob agree on a security parameter $N$;
for simplicity we take $N$ to be even.   

{\bf 2.} $A_c$ generates two independent $N$-bit random classical
strings, $S^0$ and $S^1$, and securely shares string $S^i$ with 
agent $A_i$.   

{\bf Commitment} At the designated commitment point $P$, $B_c$ gives
$A_c$ a randomly chosen size $N/2$ subset $J$ of $\{ 1 , \ldots , N \}$. 
$A_c$ immediately responds by giving $B_c$ the string 
subset $\{ S^i_j \, : i = 0, 1 \, ; \, j \in J \}$.  These string elements
are sent with their labels $i$ and $j$.  
To commit to bit value $i$, $A_c$ also sends the complementary
string subset \mbox{$S^i_{\bar{J}} = \{ S^i_j : j \in \bar{J} \}$.}  These string elements
are sent with their labels $j$, but not $i$.   Hence $B_c$ refers to 
this string as $S_{\bar{J}}$, with $i$ unknown to him.   

{\bf Unveiling} For $i \in \{0, 1 \}$: 

At the point $Q_i$, if agent $A_i$ believes Alice
wishes to unveil, she gives $B_i$ the complete string $S^i$.  

{\bf Verification}  On receiving the two substrings $S^0_J$ and $S^1_J$, $B_c$ checks 
that their Hamming distance satisfies
$ | d (S^0_J , S^1_J ) - N/4 | <  C N^{3/4} $,  
where $C>0$ is an agreed parameter independent of $N$. 
If not, he immediately rejects the commitment. 
Otherwise, Bob's agents between $P$ and $Q_i$ wait for the
data from $A_c$ and $A_i$. As soon as an agent receives
data from $B_i$ and $B_c$, he checks whether the string $S^i$ reported
by $B_i$ is the union of the strings $S^i_J$ and
$S_{\bar{J}}$ reported by $B_c$, with matching labels for each
element.   If it is,  
he accepts that Alice made a valid commitment to bit value $i$.

\vspace{2mm} {\bf Security against Alice}  $A_0$, $A_c$ and $A_1$
announce their outcomes at spacelike separation from one another
during the protocol.  They may pre-agree a collective strategy $S$, which
may rely on shared quantum or (hypothetical) post-quantum
no-signalling resources, but cannot signal to one another during the protocol.  
We are interested in bounding the probability that $A_0$ and $A_1$
both succeed in passing Bob's tests for a valid commitment of 
$0$ and $1$ respectively.   

Let $p_0 (S)$ and
$p_1 (S)$ be the probabilities that when Alice employs strategy $S$,
she and her agents convince Bob that they have have unveiled $0$ or
$1$ respectively according to the rules of the protocol above.  Note
that these rules allow for the possibility that Bob accepts valid
commitments to both $0$ and $1$ if both tests are passed.
Let $p_0 (S) + p_1 (S) = 1 + \epsilon^3_S(N)$.

Hence for any strategy $S$, $\epsilon^3_S(N) \leq \sum_{L}
\frac{p^S_{J}}{2^{N}}$ where $p^S_{J}$ is the probability
that when Alice employs strategy $S$ and Bob chooses the subset
$J$, Alice's agents produce strings $S^0$ and $S^1$ such
that 
$ | d (S^0_J , S^1_J ) - N/4 | <  C N^{3/4} $ and 
also that $S^0_j = S^1_j = S_j $ for $j \in \bar{J}$.  

If Alice's agents succeed in producing strings with these 
properties, they can infer 
that the complement $\bar{J}$ of $B_c$'s chosen subset $J$ lies within
a subset of size no more than $3N/4 + CN^{3/4}$ of $\{ 1 , \ldots , N
\}$.  By the no-signalling principle, it follows that $\epsilon^3_S(N)
\rightarrow 0$ as $N \rightarrow \infty$. 

{\bf Security against Bob} \qquad 

Alice's strings are randomly generated.
Bob thus receives no information correlated with the bit value $i$ unless
and until at least one of the $A_i$ unveils.  

Note that this security argument relies on Alice being able to
generate $2N$ unconditionally secure classical random bits per
committed bit, to share these bits between $A_c$ and the $A_i$ 
before the protocol, and to keep them securely in the agents'
laboratories before and during the protocol.   
\vspace{1mm} 

{\bf Comparison of Classical and Device-Independent Quantum
  Relativistic Bit Commitment Protocols}   

{\bf Security} The classical and device-independent quantum relativistic 
bit commitment protocols described above all have the
same configuration of agents for the two parties, and are
all unconditionally secure.   While the quantum protocols
do not require either party to trust the quantum devices
used, they still require both parties to have trusted 
classical computing and memory devices.  They thus have
no advantage over the classical protocol in this respect. 
 
{\bf Use of Classical Randomness: Bob:} The classical protocols
and the CHSH1 quantum protocol both require $B_c$ to be 
able to generate a random string that Alice cannot predict
in advance.  These strings may be generated at $P$, and 
immediately handed over to $A_c$, so they do not need
to be kept secure; nor do they need to be distributed
securely to Bob's other agents.   

The CHSH2 protocol also requires a second random string
to be generated and securely distributed and stored by $B_0$ and
$B_1$.
The CHSH3 protocols instead requires $B_0$ and $B_1$ to generate
random independent strings that Alice cannot predict in advance.  
These strings may be generated at the points  $Q_i$, and 
immediately handed over to Alice's agents $A_i$, so they do not need
to be kept secure; nor do they need to be distributed
securely to Bob's other agents.   

{\bf Use of Classical Randomness: Alice:} The classical protocol
requires Alice to generate two $N$-bit random strings, to 
share them securely between $A_c$ and $A_0$ or $A_1$,
and to store them securely in the various agents' laboratories.
In contrast, the CHSH protocols each require $A_c$ to 
generate and keep secure a single random classical bit. 
If the devices are distinguishable, this bit is also shared and stored securely
in $A_0$ and $A_1$'s laboratories, since the set of physical
devices (purported stored qubits) given to these agents 
depends on the bit value.

In addition, the CHSH protocols require $A_c$ to share
devices (purportedly containing sets of qubits) with $A_0$ and $A_1$, 
and these devices must also be stored securely in
the respective laboratories.   If Bob can access
all the devices, he can gain information about the
committed bit.

This adds nuance to the comparison. 
If the devices do indeed contain 
sets of qubits, and each qubit is indeed maximally entangled with a
another qubit in another appropriately located device, and 
the devices do indeed faithfully 
carry out projective CHSH measurements on these qubits, Alice could
use them to generate classical random strings shared 
between $A_c$ and each of the $A_i$, and these shared strings
could be used to implement the classical protocol $RCCBC$.
However, since the devices are untrusted, Alice cannot
naively implement such a strategy. Moreover, although it may be harder to keep 
entangled quantum states secure than to keep shared
random classical strings
secure, the former offers a potentially valuable
security advantage.  Imagine a scenario in which
Alice has
a large network of agents who carry out commitments,
generally relying on untrusted quantum devices.
For example, Alice may be technologically disadvantaged, and so must
generally rely on externally supplied devices, which may have been
produced by Bob.   However, suppose that, at some extra cost, Alice is able, without prior
announcement, to introduce trusted quantum devices 
to verify some of the 
purported singlets used by her agents.   She can then get information about the extent of 
any potential security breaches.  
If her agents are using the classical protocol, on the other hand,   
there is no reliable way for her to verify the 
privacy of the shared classical random strings she uses.
If Bob is able to breach her security, he can gain information
about the strings (and hence her bit commitments) without 
altering them or leaving any other physical trace.  

{\bf Summary}  Both our device-independent
CHSH quantum relativistic bit commitment
protocols and their classical counterpart
are unconditionally secure. 
However, the device-independent protocols appear to have
some subtle but real advantages that are potentially
valuable in some scenarios.  
We thus take our discussion as offering an
argument for the practical value of device-independent
quantum cryptography in relativistic
bit commitment, and stronger arguments
for its value in more general relativistic cryptographic
scenarios.   We should, though, underline that we
have compared specific device-independent protocols
against a specific classical protocol. 
It remains an open task to identify the optimal
protocols of each type (for any of the various natural measures of
optimality) and compare them. 

\acknowledgments

This work was partially supported by an FQXi mini-grant and by
Perimeter Institute for Theoretical Physics. Research at Perimeter
Institute is supported by the Government of Canada through Industry
Canada and by the Province of Ontario through the Ministry of Research
and Innovation.   AK thanks Serge Massar for helpful conversations.


\end{document}